\documentclass[conference]{IEEEtran}
\IEEEoverridecommandlockouts
\usepackage{cite}
\usepackage{amsmath,amssymb,amsfonts}
\usepackage{algorithmic}
\usepackage{graphicx}
\usepackage{textcomp}
\usepackage{xcolor}

\usepackage{amsthm}
\theoremstyle{definition}
\newtheorem{definition}{Definition}

\PassOptionsToPackage{hyphens}{url}\usepackage[]{hyperref}
\hypersetup{
    colorlinks=true,
    linkcolor=blue,
    filecolor=magenta,      
    urlcolor=cyan,
}

\makeatletter
\def\footnoterule{\kern-3\p@
  \hrule \@width 2in \kern 2.6\p@} 
\makeatother
\newcommand{\copyrightnotice}[1]{{%
  \renewcommand{\thefootnote}{}
  \footnotetext[0]{#1}%
}}

\begin{document}

\title{A Study on Shared Objects in Sui Smart Contracts

}

\author{\IEEEauthorblockN{Roman Overko}
\IEEEauthorblockA{\textit{Department of Research and Development} \\
\textit{IOTA Foundation}\\
10437 Berlin, Germany \\
roman.overko@iota.org}
}

\maketitle
\copyrightnotice{
    \textcopyright 2024 IEEE. Personal use of this material is permitted. Permission from IEEE must be obtained for all other uses, in any current or future media, including reprinting/republishing this material for advertising or promotional purposes, creating new collective works, for resale or redistribution to servers or lists, or reuse of any copyrighted component of this work in other works. 
}

\begin{abstract}
In many smart contract architectures, every contract or 
object is mutably shared by default. The Sui smart 
contract platform bears the unique feature of distinguishing between 
shared and owned objects. While transactions operating on shared 
objects require consensus to sequence reads and writes, those 
involving only owned objects are independent and may bypass consensus;
thus, the latter are less prone to this throughput bottleneck.
However, it may not always be 
possible or desirable to avoid using shared objects.
This article aims at identifying and investigating decentralized 
applications that require shared objects. Utilizing the Sui Rust 
SDK to query programmable transaction blocks, we analyze the 
frequency of transactions involving shared objects, 
shared resource contention levels, and most ``popular" 
applications that contain shared objects. The presented results are 
reproducible and show the extensive usage of shared objects in Sui, low 
contention levels, and moderate dependency among shared objects in 
atomic transactions. This novel study of shared object use cases in
a relatively new smart contract platform is important for improving the efficiency of 
such object-based architectures. This work is relevant 
for smart contract platform designers and smart contract developers.
\end{abstract}


\section{Introduction}\label{sec:introduction}

As blockchain technologies thrive, choosing the right smart 
contract platform becomes supreme for building dependable
and flexible decentralized applications (dApps). Ethereum Solidity, 
Cardano Plutus, Sui Move are examples of prominent smart 
contract frameworks, with Ethereum being the first and 
still the most popular for 
smart contracts~\cite{metcalfe2020ethereum, move}.

In Ethereum, as well as in many other smart contract platforms, every 
contract or object is public and mutably shared by default, and it can 
be thought of as an open API. This means anyone can call other smart 
contracts in their own smart contracts~\cite{eth-smart-contracts}. 
Ethereum smart contracts are not controlled by a user. Instead, 
they are deployed to the network, run as programmed, and are 
controlled by the logic of the smart contract code. All 
variables/objects on Ethereum are accessible for reading by 
everyone in the contract storage. The only way to interact with 
a smart contract is through a function call. To prevent anyone 
from calling certain functions, the contract creator needs to
implement additional access control rules. For example, the 
contract code may require the message sender to match 
the contract owner's address to allow writing to a variable, 
meaning that only the contract owner can call the function 
that mutates the variable. This technique is very similar to the 
embedded permission pattern that restricts the invocation of 
individual functions to a permissioned 
set of accounts by embedding permission controls into the 
contract~\cite{emb-permission}.

Similarly to Ethereum, Cardano smart contracts are publicly 
shared and available to everyone, unless the contract code 
includes additional access control logic. However, compared 
to Ethereum, concurrent operations on smart contracts are 
currently not possible in Cardano. Specifically, Cardano's Extended 
UTXO (EUTXO) model is limited to one state change (transaction (TX)) 
per block per smart contract~\cite{cardano-concurrency, chakravarty2020extended}. 
In other words, a shared resource (i.e., script EUTXO) 
can only be accessed once per block. Whenever a user needs 
to interact with a Cardano smart contract, they need to 
lock that contract for one block, which means that if other users 
want to interact with the same contract, only one 
interaction (TX) will succeed, as UTXOs can only 
be spent once. This bottleneck limits the throughput 
in DeFi and also complicates the development of 
dApps that require shared resources on Cardano, as smart 
contract developers have to address concurrency and contention 
issues. Various solutions to this problem have been proposed, and usually involve an off-chain third party 
(e.g., bots, batchers) that collects concurrent TXs 
and executes them as a single TX~\cite{cardano-concurrency, utxo-scaling}.

Sui is a relatively new smart contract platform (also a DLT with a DAG-based mempool~\cite{narwhal} and consensus~\cite{bullshark}) that finds a middle ground between 
the Ethereum and Cardano smart contract architectures by 
employing an object-based model and distinguishing between 
shared and owned (single-writer) objects~\cite{sui, lutris}.
In general, a smart contract developer should prefer owned 
objects to shared ones whenever it is reasonable or possible: in contrast to TXs operating on shared 
objects (which require sequencing and, thus, lead to a 
throughput bottleneck), TXs involving only owned 
objects may bypass consensus on ordering and do not need to be sequenced~\cite{fastpay, lutris}.
However, it may not always be possible to avoid shared objects. 
In many dApps, multiple users can interact with the same 
contract nearly at the same time. A DEX is an example of such a dApp: multiple 
swap TXs want to operate on the same liquidity pool, and 
it is often necessary to have a global view of all liquidity or 
total token supply to determine actions. Moreover, some DeFi concepts (e.g., constant-product AMM) may only be possible 
with shared objects; otherwise, an off-chain third party is required~\cite{dex-amm}.

This work aims at identifying and investigating dApps that require shared objects. In particular, we analyze
how often shared objects are involved in TXs on Sui and 
the level of contention and dependencies for shared objects. Our
motivation for this analysis is that understanding shared-object 
use cases is a crucial step in improving the efficiency of 
object-based smart contract platforms. Furthermore, this work may 
be helpful in designing a more flexible and fine-grained model 
for shared objects than Sui's one. Since Sui is a relatively new object-based smart contract platform, to the best of our knowledge, we believe this study on shared objects is novel.

Some of the key findings and contributions of this work are as follows:
(i) shared objects are extensively used in Sui; (ii) contention 
levels are relatively low; (iii) dependencies among shared 
objects in an atomic TX are moderate; and (iv) 
liquidity pools are a quite popular use case of shared objects.

The structure of this paper is as follows. In the next section,
we provide a detailed description of the question and describe
the methodology used in this work. Section~\ref{sec:results} is devoted 
to the presentation and interpretation of the results, a short overview of related work. In the last section, 
we conclude our findings and discuss future research directions.
\section{Methods}\label{sec:methods}
Sui is a layer-1 smart contract platform based on an object-centric 
model. The Sui ledger stores a collection of programmable objects, 
each with a globally unique identifier. The ability to distinguish 
between different kinds of objects (owned, immutable, shared)
is a unique feature of Sui~\cite{sui}.

{\it Owned objects} are the most common type of object in Sui. Many TXs and operations with assets (such as asset transfers, NFT minting, smart
contract publishing, etc.) can be designed using exclusively owned objects.
An owned object is a {\it single-writer}, meaning that only the owner
can access it via a read or write operation at a time. Since only a single
owner can access objects of this type, TXs involving exclusively owned
objects can be executed in parallel with other TXs that have no objects in
common. In other words, TXs involving only owned objects do not need
to be sequenced, and thus, they may bypass consensus in Sui.

{\it Shared objects} can be accessible to everyone for reading or writing on
the Sui network. If needed, extended functionality and accessibility of
shared objects require employing additional access logic. Some use cases
(such as the pull model of price feed updates, an auction with open bidding,
or a central limit order book) require shared objects. A shared
object is a {\it multi-writer}, meaning it can be accessed by two or more users
simultaneously. In contrast to TXs that involve only owned objects,
TXs operating on shared objects require consensus to sequence (order)
reads and writes.

Since shared objects may be required to implement the logic of 
accessibility by multiple users simultaneously, it is not 
always possible or desirable to use only owned objects.
By analyzing Sui's object-based ledger, this work aims to 
address the following questions: {\it How often are shared 
objects used in Sui? What are the use cases
for shared objects? What are the most popular
applications utilizing shared objects on Sui?  
How high are the levels of contention and dependency 
among shared objects in Sui?} In the next section, we
define some concepts and metrics used to address
these questions and describe how data was collected 
for this study.

\subsection{Terminology and Metrics}\label{sec:metrics}

\begin{definition}[Object]\label{def:object}
    An object is the basic unit of storage.
\end{definition}

\begin{definition}[Owned objects]\label{def:owned-objects}
    An owned object is an object owned by an address (or another object) and
    can only be used in TXs signed by the owner address at a time.
\end{definition}

\begin{definition}[Shared objects]\label{def:shared-objects}
    A shared object is an object that does not have a specific owner. Anyone 
    can read or write (if additional access control rules are not set) 
    shared objects.
\end{definition}

\begin{definition}[Epoch]\label{def:epoch}
    In Sui, each epoch takes $\approx24$ hours.
\end{definition}

\begin{definition}[Checkpoint]\label{def:checkpoint}
    A checkpoint (also called sequence number) in Sui changes approximately 
    every $2-3$ second.
\end{definition}

To estimate how often Sui TXs operate on shared objects 
and evaluate shared resource contention levels, new
concepts and metrics are introduced and defined as follows.

\begin{definition}[Interval]\label{def:interval}
     An interval is a period of time expressed in the number of checkpoints.
\end{definition}

\begin{definition}[Contention]\label{def:contention}
    Contention is a situation when multiple TXs touch 
    the same shared object at the same time, i.e., concurrently 
    access that shared object.
\end{definition}

\begin{definition}[Shared-object transaction]\label{def:sh-obj-tx}
    A shared-object TX has at least one shared
    object in its inputs.
\end{definition}

\begin{definition}[Density]\label{def:density}
    The density is the ratio of the number of shared-object TXs to the number of all TXs. The density is a number between $0$ and $1$; the higher the density, the more TXs operate on shared objects.
\end{definition}

\begin{definition}[Contention degree]\label{def:contention-degree}
    The contention degree is the ratio of the number of shared-object
    TXs (within some interval) to the number of shared objects touched by those 
    TXs (within the same interval). The contention degree is a number between $0$ and $\infty$. A contention degree of $1$ means that each shared-object TX operates on a single different object, on average; values larger than $1$ indicate multiple shared-object TXs contending for the same shared object; values smaller than $1$ mean a TX touches multiple shared objects, on average.
\end{definition}

\begin{definition}[Contended fraction]\label{def:contended-fraction}
    The contended fraction is the ratio of the number of shared
    objects (within some interval) touched by more than one TX to the total number of
    shared objects (within the same interval). The contended fraction is a number between $0$ and $1$. The higher the contended fraction, the more shared objects are touched by more than one TX.
\end{definition}

\subsection{Data collection}\label{sec:data-collection}
For this analysis, we used the Sui Rust SDK~\cite{sui-rust-sdk} to query 
all Sui programmable TX blocks starting from epoch $0$ 
(April 12, 2023), i.e., the genesis, until epoch 
$315$ (February 22, 2024), inclusively, which resulted in 
a total of $1,103,018,902$ TXs on the Sui mainnet. The source code 
for this analysis is publicly available on GitHub~\cite{project-repo}.

{\it Bullshark Quests:}
Before proceeding to the interpretation of the results in the next section,
it is worth mentioning Sui's {\it Bullshark Quests} ({\it BQs}), which have been an 
ongoing initiative offering the opportunity to earn 
rewards by engaging with applications on Sui~\cite{bullshark-aces}. Each 
BQ is announced and launched by for a specific 
period of time. As we will see later in the next section, the metrics defined in 
Section~\ref{sec:metrics} may be significantly different during the 
periods of BQs compared to those during periods when 
the quests did not take place. The time frames of BQs until epoch $315$ are as follows:
\begin{itemize}
    \item {\it BQ-1} started on July 6, 2023 (epoch $85$)
        and ended on July 27, 2023 (epoch $106$)~\cite{bullshark-quest-1}.
    \item {\it BQ-2} started on July 28, 2023 (epoch $107$) 
        and ended on September 5, 2023 (epoch $146$)~\cite{bullshark-quest-2}.
    \item {\it BQ-3} started on October 12, 2023 (epoch $183$) 
        and ended on November 9, 2023 (epoch $211$)~\cite{bullshark-quest-3}.
    \item {\it Winter Quest} ({\it WQ}) started on December 18, 2023 (epoch $250$) 
        and ended on December 26, 2023 (epoch $258$), when all rewards were 
        claimed~\cite{winter-quest}.
\end{itemize}
\section{Results and Discussion}\label{sec:results}
In this section, we estimate how often Sui TXs operate on shared objects, evaluate shared resource contention 
levels using metrics defined in Section~\ref{sec:metrics}, discuss the results, and give a brief overview of related work.

\subsection{Number of transactions}\label{sec:tx-number}
Even though the number of TXs per epoch is not of a particular
interest in this analysis, we begin by presenting this metric as it
is used to calculate and interpret the density.

\begin{figure}[ht]
    \centering
    \includegraphics[width=0.95\columnwidth]{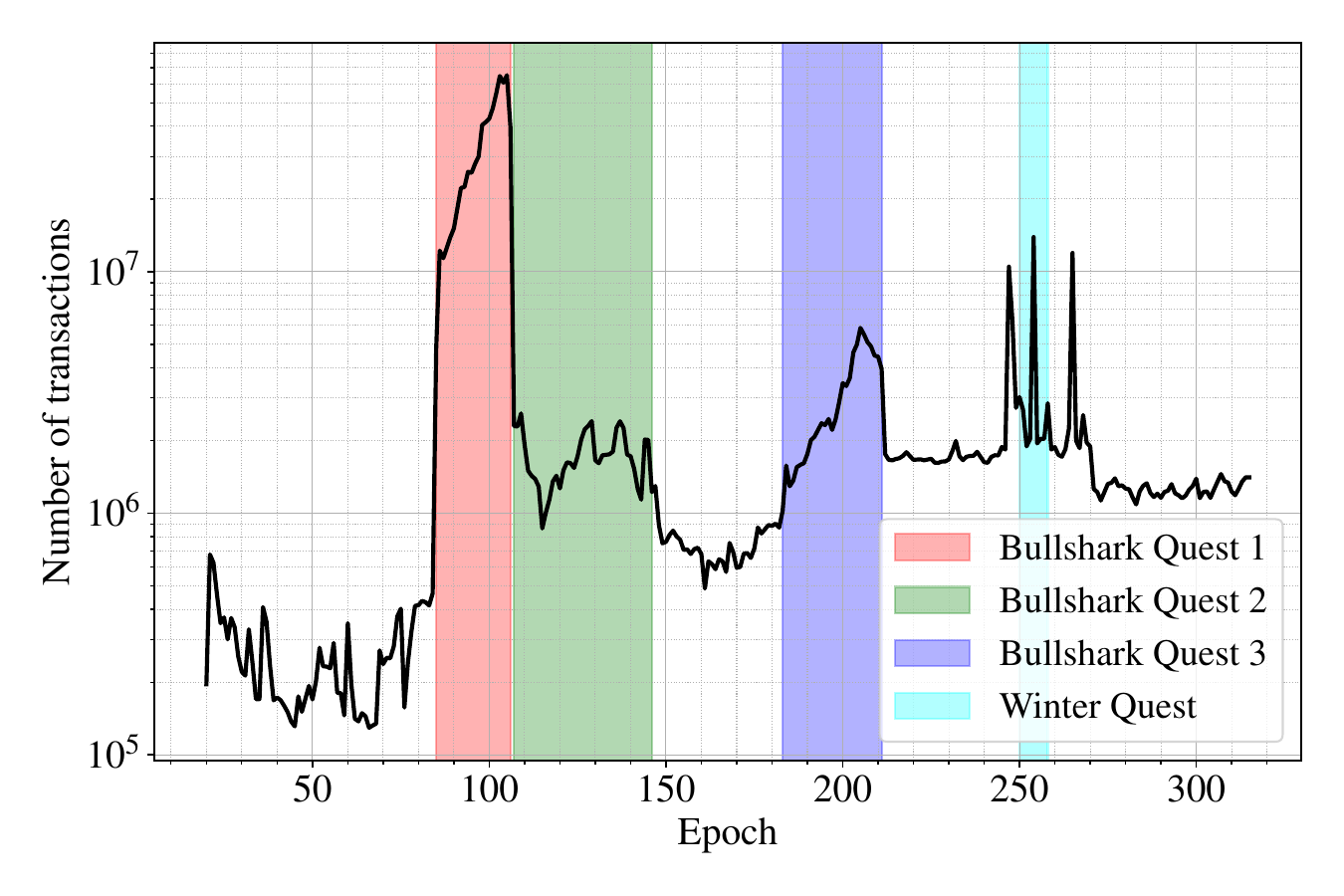}
    \caption{Number of TXs per epoch in Sui. Note a log scale 
    on the y-axis.}
    \label{fig:tx-number}
\end{figure}

Figure~\ref{fig:tx-number} shows the total number of TXs (on a log scale graph) processed by the Sui mainnet per epoch.\footnote{It is 
worth noting that in this and the following figures we plot the metrics 
starting from epoch $20$ since the Sui network was in a bootstrapping 
phase during those early epochs and there were almost no shared objects involved.}
As it can be seen, the number of 
TXs per epoch generally increased during the periods of 
BQs (depicted as vertical spans of different colors)
compared to their number at epochs during which BQs did not 
take place. A significant increase in the number of TXs per
epoch can be observed during the period of BQ-1 (vertical
light red span). Overall, the number of TXs per 
epoch was slightly higher than one million during periods 
without quests for epochs after BQ-3 ended.

\subsection{Density of shared-object transactions}\label{sec:density}
\begin{figure}[ht]
    \centering
    \includegraphics[width=0.95\columnwidth]{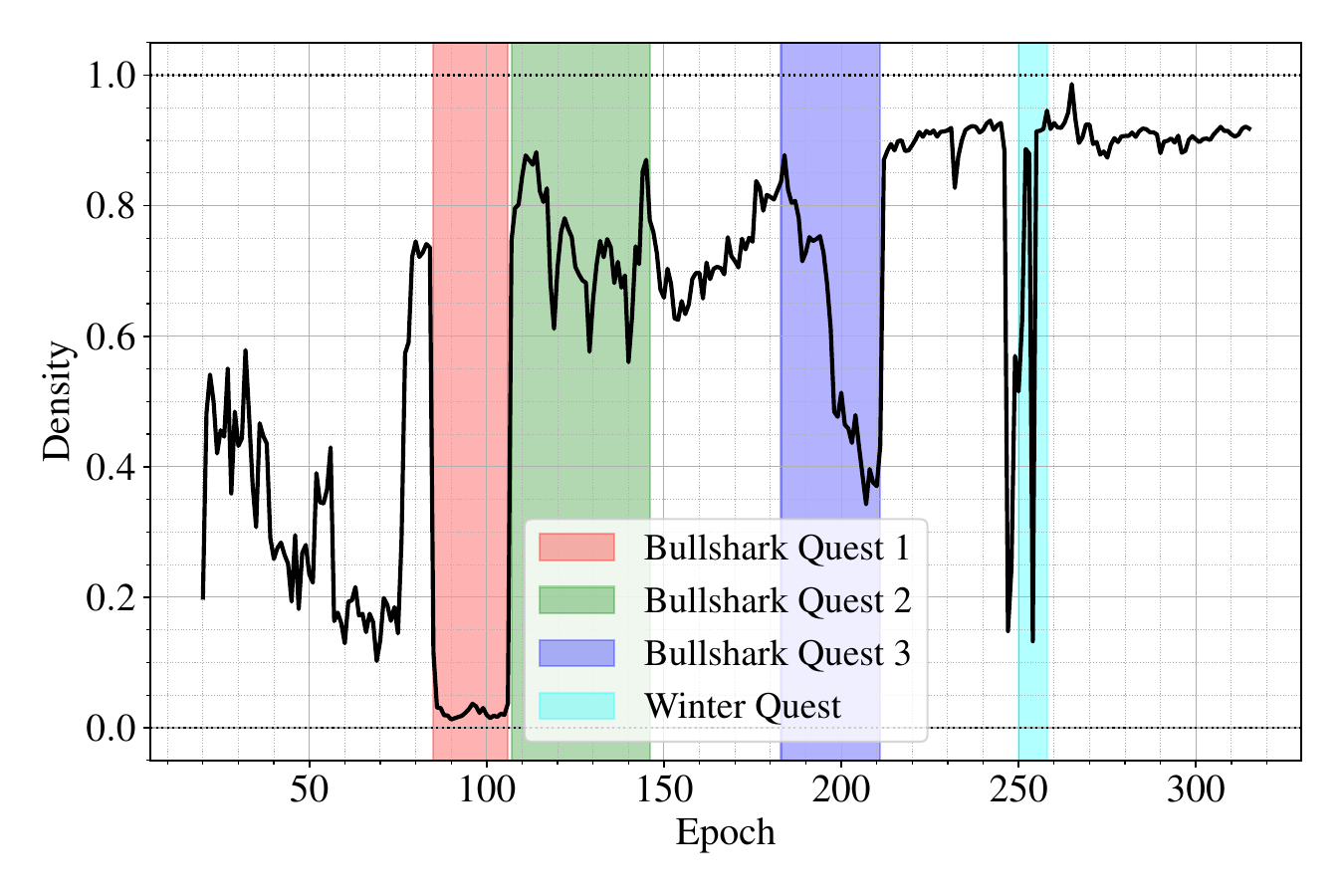}
    \caption{Density of shared-object TXs per epoch in Sui.}
    \label{fig:density}
\end{figure}

Recall from Definition~\ref{def:density} that the density is defined
as the ratio of the number of shared-object TXs to the total 
number of TXs within some time interval.
Figure~\ref{fig:density} depicts the density of shared-object
TXs per epoch.

As it can be seen in Figure~\ref{fig:density}, Sui TXs extensively
operate on shared objects overall, especially starting from epoch $107$ 
(BQ-2 start). At the beginning, starting from epoch $20$,
the density increased to approximately $0.5$ and fluctuated around
this value until epoch $35$---during this period, many frequently 
used shared-object applications, such as \texttt{Sui 
Framework}~\cite{sui-framework}, \texttt{Pyth Network}~\cite{pyth-network}, 
\texttt{Cetus}~\cite{pyth-network}, \texttt{Kriya DEX}~\cite{kriya}, 
\texttt{Turbos Finance}~\cite{turbos}, were deployed on Sui~\cite{project-repo}.
After this period, a trend of decreasing density to a value of $0.2$ can be 
observed until epoch $80$, after which it peaked at $\approx0.75$. 
This peak can be explained by the deployment of new frequently used 
shared-object applications, such as \texttt{DeepBook}~\cite{deepbook} 
and \texttt{DeSuiLabs Coin Flip}~\cite{desuilabs}.

During BQ-1 (i.e., from epoch $85$ until $106$), the density 
dramatically decreased to very low values ($<0.05$). To better 
understand this decline, we refer to Figure~\ref{fig:tx-number}, which 
shows a significant rise in the number of TXs per
epoch during the BQ-1 period. Taking this observation and 
low values of density into account, it can be concluded that TXs 
related to BQ-1 did not use shared objects extensively.

For BQ-2, it can be observed that the corresponding period is 
characterized by the number of TXs per epoch being smaller by 
one order of magnitude than that for the BQ-1 period 
(see Figure~\ref{fig:tx-number}). Despite this, the density dramatically 
increased once BQ-2 started and remained high (fluctuating
around $0.7$) until BQ-3 (see Figure~\ref{fig:density}), which 
indicates that TXs in BQ-2 extensively operated on shared objects---indeed, BQ2 participants could earn rewards by engaging with more shared-objects dApps (e.g., Cetus~\cite{cetus}, Kriya~\cite{kriya}, Scallop~\cite{scallop}, Turbos~\cite{turbos}) than in BQ1~\cite{bullshark-quest-1, bullshark-quest-2}.

Starting from the BQ-3 start, the density gradually
decreased to approximately $0.4$ over the corresponding period. After
BQ-3 ended, the density of shared-object TXs
reached extremely high values and fluctuated around $0.9$, except for
a few epochs before WQ and during its period when it dropped 
below $0.2$. As can be concluded, shared objects are, in general, 
extensively used in Sui, especially 
when the network becomes more mature (from epoch $211$ until epoch 
$315$, see Figure~\ref{fig:density}).

\subsection{Contention degree}\label{sec:contention-degree}
The density is quite a simple metric that provides an overall picture of
how often shared objects are involved in TXs. However, this
metric does not capture information about how many TXs 
{\it contend for} (operate on) the same shared object within some time interval.
Consider one scenario with many shared-object TXs, each operating 
on a different shared object, and another scenario when the same number of 
shared-object TXs operate on a few shared objects
(assume the total number of TXs in both cases is equal). 
In both scenarios, the density of shared-object TXs would be the 
same, even though these two cases are moderately different. Since shared objects 
are usually multi-writers, multiple TXs may contend for the same 
object (the second scenario), which requires sequencing for execution.
However, in the first scenario, there is no contention as each TX operates on a different shared object, and thus, they are independent
and can be executed in parallel. To capture such important details in
our analysis, we use another metric called {\it contention degree}, 
as defined in Definition~\ref{def:contention-degree}.

\begin{figure}[ht]
    \centering
    \includegraphics[width=0.95\columnwidth]{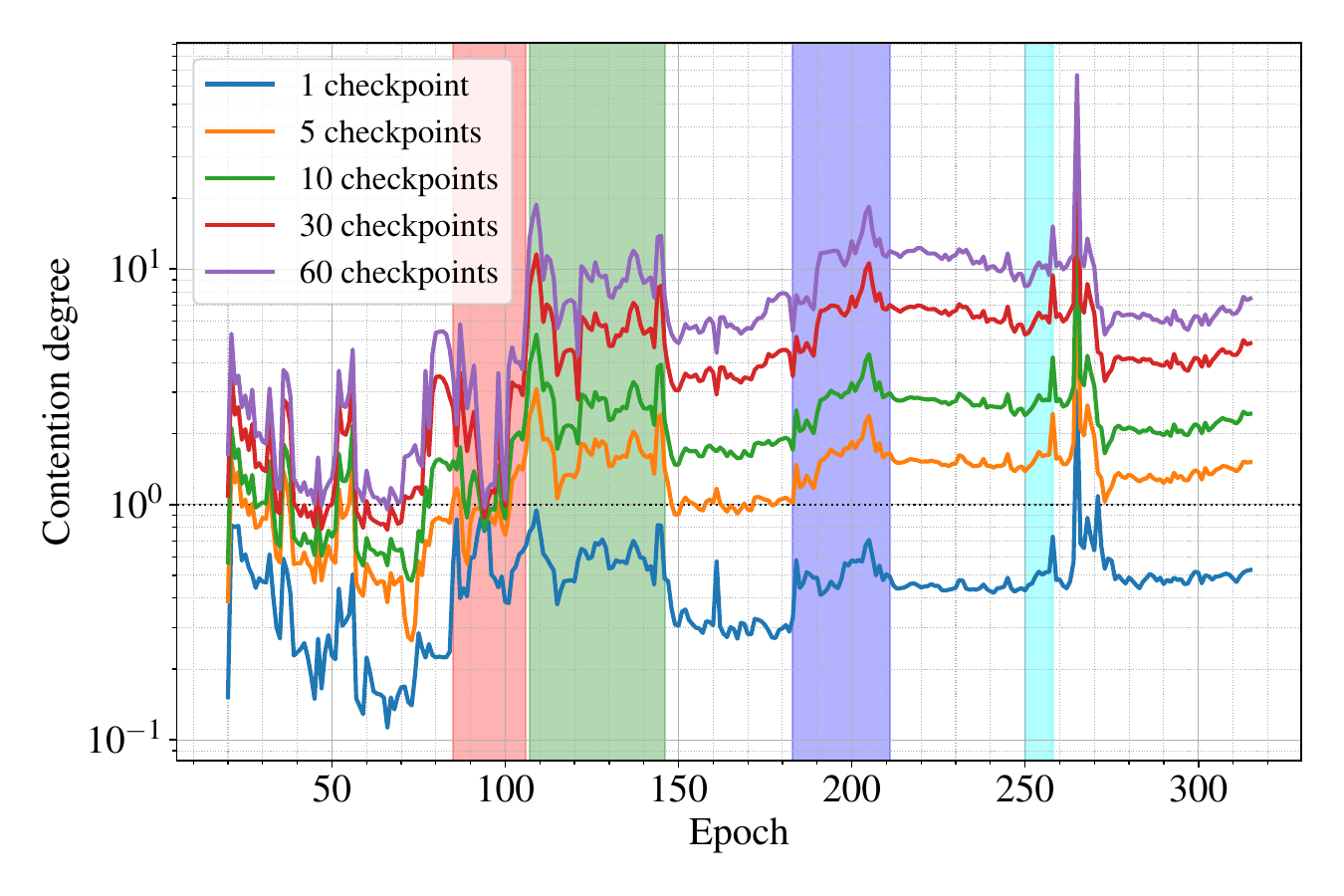}
    \caption{Averaged contention degree per epoch for different time intervals
        in Sui. Note a log scale on y-axis.}
    \label{fig:contention}
\end{figure}

Figure~\ref{fig:contention} shows the contention degree averaged
over time intervals of various lengths, per epoch. That is, for a given time interval 
(see Definition~\ref{def:interval}), we count the number of shared-object 
TXs in the interval and divide that number by the number of shared 
objects used in the same interval. Such ratios are then summed up, and the 
sum is divided by the number of intervals in an epoch (see 
Definition~\ref{def:epoch}). Since the sequencing of shared-object TXs 
in Sui happens on a per-checkpoint basis, the shortest time 
interval in our analysis equals one checkpoint (see Definition~\ref{def:checkpoint}).
Intervals of various lengths are used to investigate how contention levels
change with increasing interval duration.

From Figure~\ref{fig:contention}, it can be seen that the average contention
degree for the interval of one checkpoint is less than $1$ almost across all epochs, meaning that each shared-object TX operates on a single different object on average. For intervals longer than one checkpoint, we can observe a significant rise 
in the average contention degree after BQ-1 ended, which indicates 
high involvement of shared objects in TXs when the network becomes mature. 
Finally, as expected, the longer the interval,
the higher the contention degree, which implies that contention for shared
objects is likely to be less prominent in ``fast-committing'' consensus protocols.

\subsection{Contended fraction}\label{sec:contended-fraction}
\begin{figure}[ht]
    \centering
    \includegraphics[width=0.95\columnwidth]{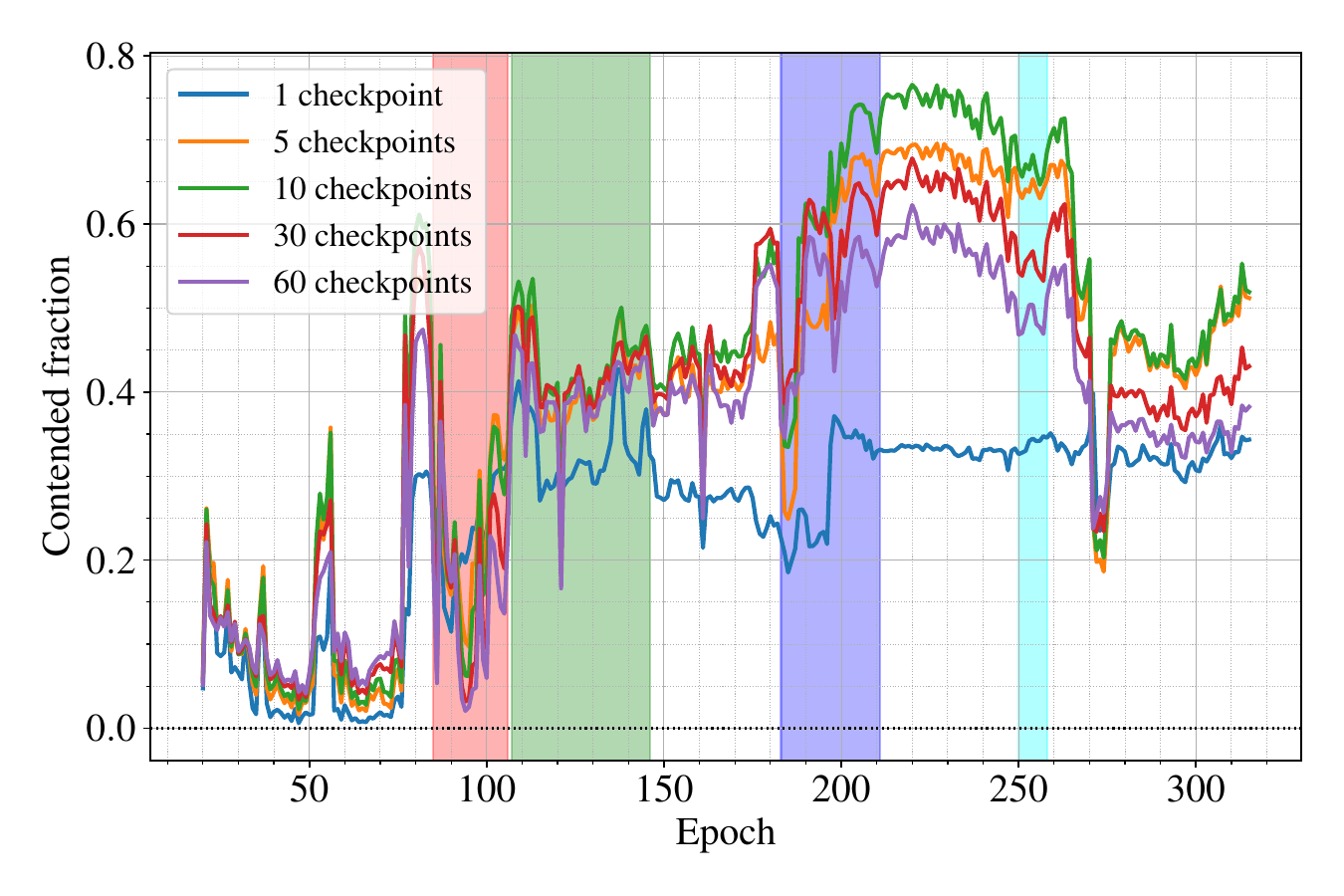}
    \caption{Averaged contended fraction per epoch for different time intervals.}
    \label{fig:contended-fraction}
\end{figure}

To further investigate contention for shared objects in Sui, we use a
metric called {\it contended fraction} (see Definition~\ref{def:contended-fraction}), 
which gives the frequency of shared objects touched by more than one TX 
within some time interval. Figure~\ref{fig:contended-fraction} depicts the 
contended fraction averaged over intervals of various lengths, per epoch. That is,
for a given interval, we count the number of shared objects touched by more
than one TX in that interval and divide it by the number of
shared objects. Such ratios are then summed up, and the sum is divided
by the number of intervals in an epoch.

A few conclusions can be drawn from Figure~\ref{fig:contended-fraction}. First,
the averaged contended fraction does not vary much with increasing 
duration of an interval at the early stages of the network. Second, more 
important, shared objects are more frequently touched by only one 
TX rather than multiple TXs on average: for an 
interval of one checkpoint (the actual frequency of commitments in Sui),
the average contended fraction fluctuates around $0.3$, which can be 
easier observed for epochs when the network is more mature (e.g., after 
BQ-3 ended). Third, the average contended fraction
greatly increases when the duration of the interval is lengthened for
a more mature network, as can be seen in the figure after BQ-2 ended.
Finally, longer intervals do not necessarily imply
larger values of the average contended fraction.

\subsection{Number of shared objects in transaction 
    inputs}\label{sec:shared-obj-number}
\begin{figure}[ht]
    \centering
    \includegraphics[width=0.95\columnwidth]{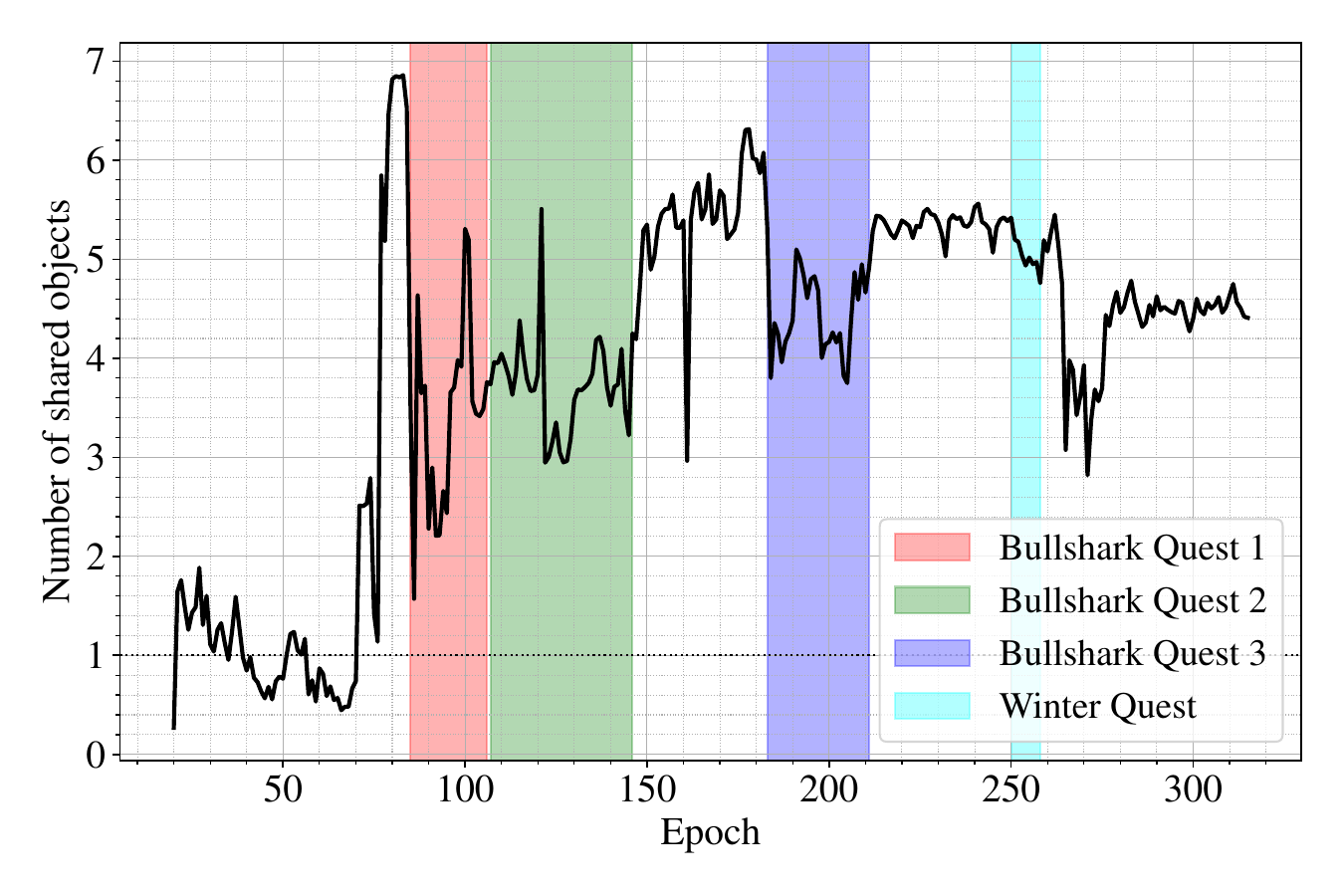}
    \caption{Average number of shared objects touched by TXs.}
    \label{fig:obj-number}
\end{figure}

While the contended fraction (see Figure~\ref{fig:contended-fraction})
shows how often shared objects are touched by more than one TX,
it does not capture information about how many shared objects are
touched in an atomic TX. Such information provides insights on composability
and dependencies in smart contracts involving shared objects. 
Figure~\ref{fig:obj-number} depicts the average number of shared objects
in TX inputs, per epoch.

As can be observed in Figure~\ref{fig:obj-number}, the average number
of shared objects in TX inputs fluctuated around $1$ when the 
network was immature, until nearly epoch $70$, when it dramatically
increased to almost $7$ shared objects per TX on
average right before BQ-1. After BQ-2 ended, it can be seen that the average number
of shared objects in an atomic TX fluctuates around a
value of $5$, which indicates a moderate degree of dependencies among
shared objects in Sui.

Figures~\ref{fig:density}-\ref{fig:obj-number} illustrate metrics
calculated based on the aggregation of all shared objects. That is, 
these metrics do not provide any information about which shared objects 
and dApps are used mostly. Sections~\ref{sec:apps} 
and~\ref{sec:shared-obj-types} address this.

\subsection{Popular shared-object applications}\label{sec:apps}
\begin{figure}[ht]
    \centering
    \fbox{\includegraphics[width=0.85\columnwidth]{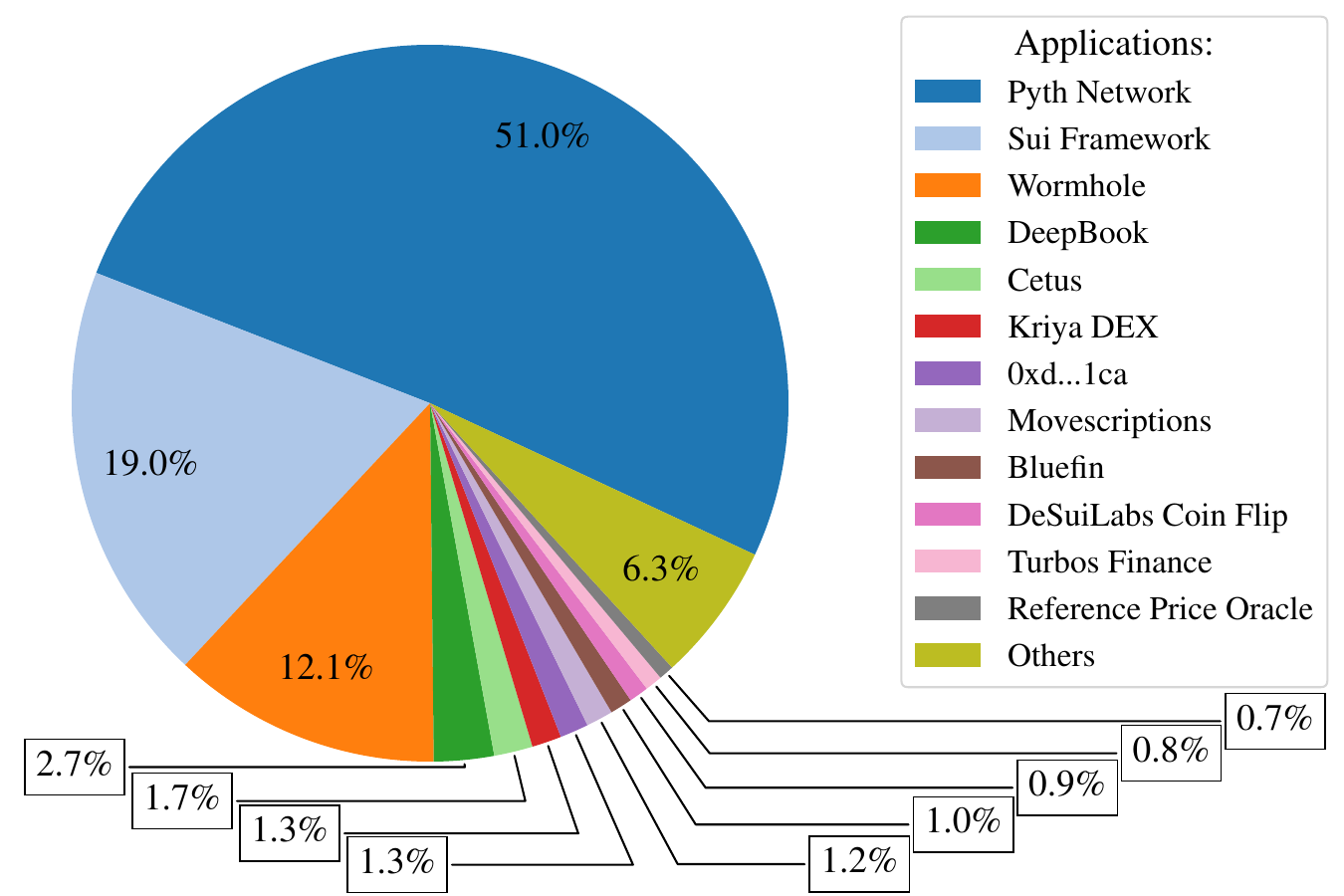}}
    \caption{The most frequently used applications involving shared objects
        in Sui.}
    \label{fig:app-pie}
\end{figure}

Before we delve into the description of particular shared objects, 
we briefly describe some of the most used applications in Sui,
as annotated in Figure~\ref{fig:app-pie}.

\texttt{Pyth Network} is a data oracle that publishes 
financial market data to multiple blockchains~\cite{pyth-network}.
Pyth price feeds employ a pull price update model, which requires
shared objects. In this model, users are responsible for 
posting price updates on-chain. \texttt{Pyth Network} 
implements two shared objects, \texttt{PriceInfoObject} and 
\texttt{State} (see Figure~\ref{fig:obj-pie}), both extensively 
used in Sui: as can be seen in Figure~\ref{fig:app-pie}, more 
than half ($51\%$) of the total number of shared-object 
TXs is taken by those related to \texttt{Pyth Network}.

\texttt{Sui Framework} provides a collection of the core 
on-chain libraries for Move developers, and it is the second most 
frequently used ($19\%$ of all shared-object TXs) 
contract involving shared objects such as \texttt{Clock}, 
\texttt{Kiosk} (both extensively used; see Figure~\ref{fig:obj-pie}),
\texttt{Table}, and others~\cite{sui-framework}.

\texttt{Wormhole} is an interoperability protocol 
powering the seamless transfer of value and information across 
multiple blockchains~\cite{wormhole}. It sends messages 
cross-chain using various verification methods to attest to the 
validity of a message. The \texttt{State} object is the only shared 
object (extensively used in Sui; see Figure~\ref{fig:obj-pie})
in \texttt{Wormhole}, and it is used (i) as a container 
for all state variables and (ii) to perform anything 
that requires access to data that defines the contract.

\texttt{DeepBook} is a decentralized central limit 
order book built for Sui to provide 
a one-stop shop for trading digital assets and to 
accelerate the development of financial and other apps 
on Sui~\cite{deepbook}. \texttt{Pool} is the only shared object type 
in the contract.





\texttt{DeSuiLabs Coin Flip} is a smart contract game for 
players to double their SUI by guessing heads or tails~\cite{desuilabs}. 
Each user's guess is represented by the \texttt{Game} shared object. 
If a guess is incorrect, the contract sends the player's 
bet into the house wallet, represented by the \texttt{HouseData} shared object.



\subsection{Popular shared object types}\label{sec:shared-obj-types}
\begin{figure}[ht]
    \centering
    \fbox{\includegraphics[width=0.85\columnwidth]{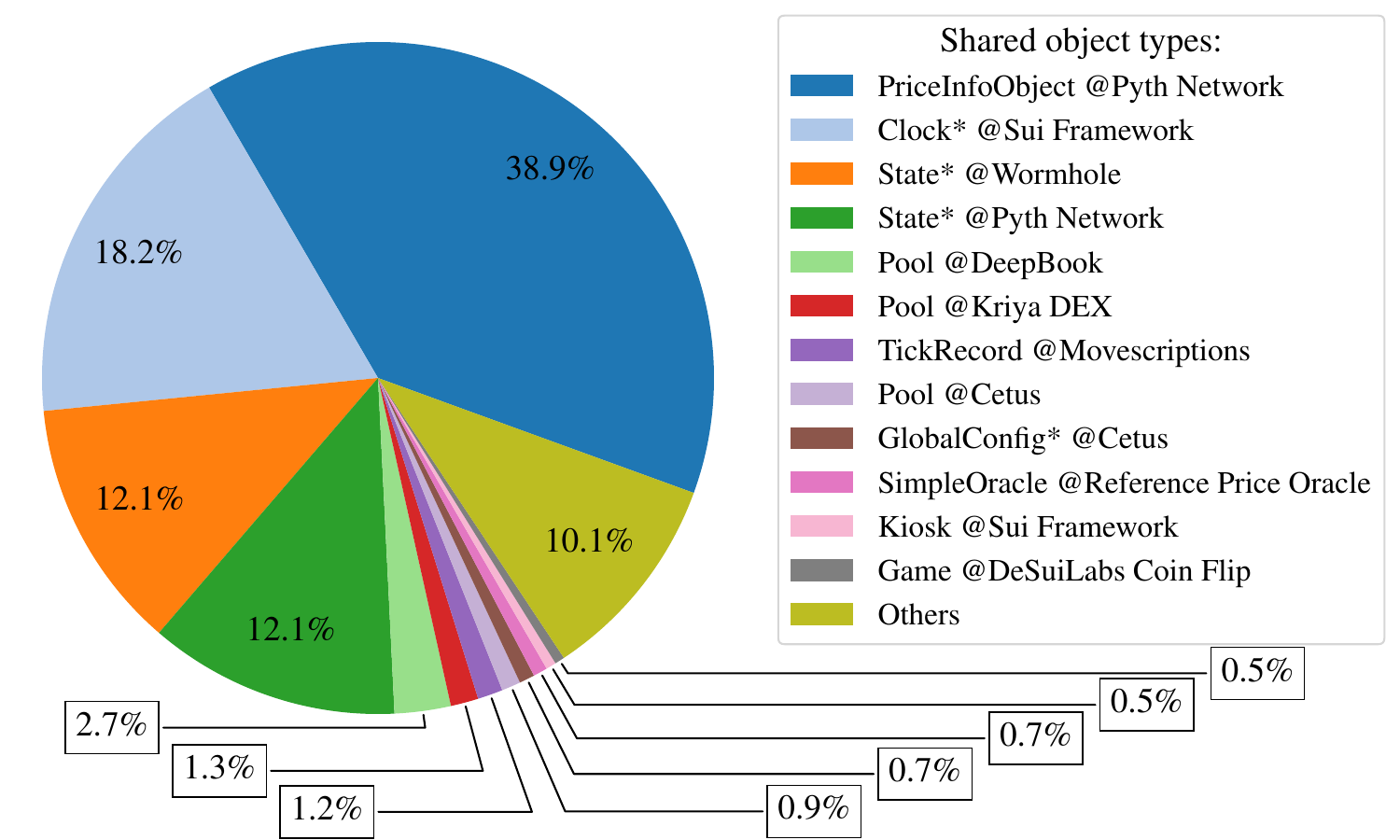}}
    \caption{Most frequently used shared object types in Sui.
    	Asterisks in the legend annotate singleton shared objects.}
    \label{fig:obj-pie}
\end{figure}

We are now in a position to describe some of the most frequently used shared
object types in Sui, as depicted in Figure~\ref{fig:obj-pie}, in which
the percentages represent a relative number of shared-object TXs 
involving those types. Asterisks in the legend annotate {\it singletons},
i.e., only one instance of such object type can be created. 
As it can be seen, the only two shared objects of \texttt{Pyth Network},
\texttt{PriceInfoObject} and \texttt{State}, constitute more than half
($51\%$) of all shared-object TXs. \texttt{PriceInfoObject} 
and some other shared objects of interest are described in detail below.

\texttt{PriceInfoObject} represents Pyth price feeds in Sui.
There are 83 instances of the \texttt{PriceInfoObject} type, 
each corresponding to a single Pyth price feed in the global storage.
TXs involving \texttt{PriceInfoObject} shared objects (i) 
constitute $38.9\%$ of all shared-object TXs (see 
Figure~\ref{fig:obj-pie}), and (ii) usually take such objects by a 
mutable reference, which likely indicates high contention for these 
shared objects~\cite{repo-epoch-315}. Both observations can be explained 
by the underlying price update model employed by Pyth, as described in
Section~\ref{sec:apps}. Typically, users of Pyth price feeds submit a 
single TX that simultaneously updates the price and uses it 
in a downstream dApp. That is, updating on-chain 
prices in Pyth is a permissionless operation, which requires employing 
shared objects.

\texttt{Clock} is a singleton shared object created during genesis,
and it is used for accessing time from Move calls using APIs provided in \texttt{Sui Framework}. Anyone can access 
\texttt{Clock} in TXs, but only via an immutable reference.
The \texttt{Clock}'s timestamp is set automatically by a system 
TX every time consensus commits a checkpoint.
Therefore, TXs that read \texttt{Clock} do not need 
to be sequenced relative to each other. However, any 
TX that requires access to \texttt{Clock} must go 
through consensus in Sui because the only available instance 
is a shared object. TXs involving \texttt{Clock} 
constitute $18.2\%$ of all shared-object TXs 
(see Figure~\ref{fig:obj-pie}).

Discussed above, \texttt{PriceInfoObject} does not have any access 
control logic: anyone can read or write it. A striking example 
of a shared object with the ownership notion is the 
\texttt{Kiosk}~\cite{kiosk}, which allows storing and trading 
any type of asset as long as the creator of those assets 
implements a transfer policy. \texttt{Kiosk} provides guarantees 
of ``true ownership'': similarly to owned objects, assets stored 
in \texttt{Kiosk} can only be managed by the kiosk owner, who can 
place, take, list items, and perform any other actions on assets
in the kiosk. Anyone can create \texttt{Kiosk}: there is a high 
number of shared objects ($428,804$) of this type~\cite{repo-epoch-315}.
By default, a \texttt{Kiosk} instance is made shared, in which case, 
the owner can sell any asset that has shared \texttt{TransferPolicy} 
available, guaranteeing creators that every transfer must be approved.
Anyone can purchase openly listed items in \texttt{Kiosk}---this 
is the only kind of write operation on \texttt{Kiosk} that can be 
performed by anyone. While purchasing the \texttt{Kiosk} items is 
available for everyone, it is possible for the owner to make a 
\texttt{Kiosk} instance an owned object. However, such a kiosk 
might not function as intended or be inaccessible to other users.

In addition to \texttt{Clock}, \texttt{Sui Framework} also provides 
other shared object types, such as \texttt{TransferPolicy}---a 
highly customizable primitive that provides an interface for the 
owner to set custom transfer rules, \texttt{TreasuryCap}---a  
capability that allows the bearer to mint and burn coins of some type, 
guaranteeing full ownership over the currency, and \texttt{CoinMetadata}---a
container for the metadata of any coin type created in Sui. While 
instances of these three shared object types are not extensively used 
in Sui TXs, it is worth mentioning an important observation: owned 
and/or immutable objects of these three types also exist~\cite{repo-epoch-315}.

A quite ``popular'' use case of shared objects in Sui is liquidity pools used in
various DEX applications and usually represented by a \texttt{Pool}
type, as in the following contracts: \texttt{DeepBook}~\cite{deepbook},
\texttt{Cetus}~\cite{cetus}, \texttt{Kriya DEX}~\cite{kriya}, 
\texttt{Turbo Finance}~\cite{turbos}, and others (see Figure~\ref{fig:obj-pie}).
A striking common characteristic of all the aforementioned \texttt{Pool} shared
objects is that they are (almost) always accessed via a write operation, which
implies high contention for these objects in TXs. It is also worth mentioning that contracts such as \texttt{DeSuiLabs Coin Flip}~\cite{desuilabs},
\texttt{Sui Framework}~\cite{sui-framework}, and \texttt{Scallop}~\cite{scallop} 
implement shared object types for which many instances have been created:
\texttt{Game} (2,702,164 instances), \texttt{Kiosk} (428,804 
instances), and \texttt{Obligation} (86,969 instances), respectively~\cite{repo-epoch-315}.


\subsection{Related work}\label{sec:related-work}
Dependency tracking in smart contracts, especially in the Ethereum ecosystem, has been widely investigated. We refer the reader to~\cite{EDIT} and references therein for a comparison of smart contract analysis tools and a study on different interaction patterns between smart contracts, externally owned accounts, and internal TXs in Ethereum. It is worth mentioning that our analysis in based on data collected from the Sui blockchain, a relatively new object-based smart contract platform. We additionally note that the topic of shared and owned objects has not gained much research attention yet.
\section{Conclusion}\label{sec:conclusion}
In this work, we analyzed shared objects on the Sui
smart contract platform. The presented results 
show that shared objects are extensively used on Sui,
especially when the network becomes more mature.
Contention for shared objects was also investigated,
and it appears to be relatively low in Sui. The most
frequently used shared objects are those related to
price feed updates and accessing time on chain.
Liquidity pools are another popular use case for shared 
objects in Sui. The average number of shared objects 
in TX inputs fluctuates around $5$, which 
indicates a moderate degree of dependencies among 
shared objects. The question of composability and 
whether those shared objects come from the same or 
different smart contracts is not present in this study but is
considered for future work, along with the evaluation of
contention levels while decoupling read and write operations. Moreover, it would be meaningful and interesting to repeat this analysis in $1-2$ years, since Sui is a relatively new and still ``young'' object-oriented smart contract platform.

\section*{Acknowledgment}
We would like to thank our colleagues at the IOTA Foundation, Dr. Can Umut Ileri, Dr. Mirko Zichichi, and Dr. Sebastian Mueller, for providing the invaluable feedback on this analysis.

\bibliographystyle{IEEEtran}
\bibliography{ref}

\end{document}